\documentclass[prl,a4paper,twocolumn,superscriptaddress,amsmath,amssymb,preprintnumbers,showpacs,nofootinbib,showkeys,tightenlines,floatfix]{revtex4}
\usepackage{latexsym}
\usepackage{epsfig}
\usepackage{amssymb}
\usepackage{amsmath}
\usepackage{color}
\usepackage{graphicx}

\newcommand{\bea}{\begin{eqnarray}}
\newcommand{\eea}{\end{eqnarray}}

\newcommand{\be}{\begin{equation}}
\newcommand{\ee}{\end{equation}}

\definecolor{Mygreen}{cmyk}{0,1.0,0,0}
\definecolor{MyOrange}{cmyk}{0.5,1.0,0,0}

\newcommand{\bi}{\begin{itemize}}
\newcommand{\ei}{\end{itemize}}

\begin{document}
\title{A trajectory approach to two-state kinetics of single particles on sculpted energy landscapes}
\author{David Wu}
\affiliation{California Institute of Technology, Pasadena CA}
\author{Kingshuk Ghosh}
\affiliation{University of California, San Francisco CA}
\author{Mandar Inamdar}
\affiliation{Indian Institute of Technology, Mumbai, India}
\author{Heun Jin Lee}
\affiliation{California Institute of Technology, Pasadena CA}
\author{Scott Fraser}
\affiliation{California Institute of Technology, Pasadena CA}
\author{Ken Dill}
\affiliation{University of California, San Francisco CA}
\author{Rob Phillips}
\email{phillips@pboc.caltech.edu}
\affiliation{California Institute of Technology, Pasadena CA}

\date{\today}

\begin{abstract}
We study the trajectories of a single colloidal particle as it hops between two energy wells $A$ and $B$, which are sculpted using adjacent optical traps by controlling their respective power levels and separation.  Whereas the dynamical behaviors of such systems are often treated by master-equation methods that focus on \textit{particles} as actors, we analyze them here instead using a \textit{trajectory}-based variational method called Maximum Caliber, which utilizes a dynamical partition function.  We show that the Caliber strategy accurately predicts the full dynamics that we observe in the experiments: from the observed averages, it predicts second and third moments and covariances, with no free parameters.  The covariances are the dynamical equivalents of Maxwell-like equilibrium reciprocal relations and Onsager-like dynamical relations.  In short, this work describes an experimental model system for exploring full trajectory distributions in one-particle two-state systems, and it validates the Caliber approach as a useful way to understand trajectory-based dynamical distribution functions in this system.
\end{abstract}

\maketitle

We explore the kinetics of two-state processes, $A \leftrightharpoons B$, at the one-particle level. Examples of single-molecule or single-particle dynamical processes that mimic this two-state dynamics include DNA loop formation  \cite{Finzi:Gelles}
, RNA oligomer hairpin formation/destruction \cite{Bustamante:Liphardt2},
protein folding oscillations \cite{Chiroco:Baldini},
sequence-dependent protein unfolding \cite{Bustamante:Collin}, or ion-channel opening and closing kinetics \cite{Auerbach:Popescu}.  Two-state fluctuating systems with constant rates are called \textit{random telegraph} processes. 

One way to understand two-state and random-telegraph processes is through master equations, which are differential equations that are solved for time-dependent probability density functions \cite{Gardiner:Stochastic}.
For single-particle and few-particle systems, however, the most direct and convenient experimental observables are often the individual dynamical trajectories, rather than the density functions.  Here, we describe an experimental model system to study single-particle two-state stochastic trajectories.  We use these experiments to test a theoretical strategy, called Maximum Caliber, that provides a way to predict the full trajectory distributions, given certain observed mean values.

Using dual optical traps, we have sculpted various energy landscapes.  We can control the relative time the particle spends in its two states and the rate of transitioning between them.  Our method follows from earlier works on the dual trapping of colloidal particles that was used to study Kramers reaction rate theory \cite{Kramers:1}.
While these experimental models were previously applied to studying average rates, our interest here is in the probability distribution of trajectories. 

We trap a 1 $\mu$m silica bead in a neighboring pair of optical traps.  The laser at 532 nm, 100 mW, provides an inverted double-Gaussian shaped potential: an acousto-optic deflector alternately sets up two traps close together in space, at a switching rate of 10 kHz, which is much faster than each individual trap's corner frequency \cite{Block:Svoboda}, and the fastest bead hopping rate. The strength of each trap and the spacing between them can be controlled in order to sculpt the shape of the potential. A tracking 658 nm red laser at 1 mW was used to determine the position of the bead.  The red laser reduces by only a small degree the trapping efficiency of the green laser. The forward scattered light is imaged through a microscope condenser onto a position-sensitive detector \cite{Block:Lang}.  The green trapping laser light at the detector is filtered out by an interference filter that passes only red light. The data was recorded at a rate of 20 kHz, which sets the fundamental time step, $\Delta t$, for our analysis.  Trajectories were recorded for intervals ranging from 20 minutes to more than 1 hour, depending on the hopping rate.  A simple threshold was used to determine states in the trajectories. 

\begin{figure}[h]
  \centering
  \includegraphics[width=6cm]{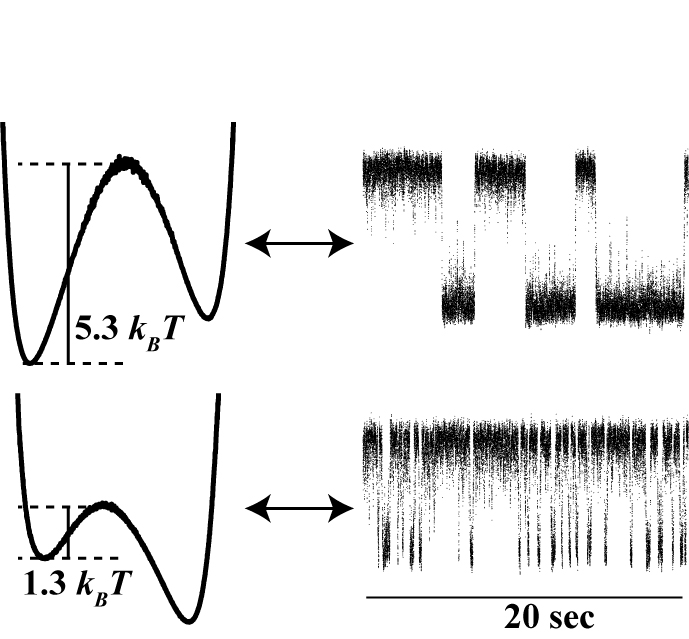}
  \caption{ \textbf{Sculpted energy landscapes (left, averaged 20 minutes) and the corresponding microtrajectories.}  The trace is raw data; states are assigned after boxcar filtering and threshold finding. Top: the lower state is slightly more populated; there is a high barrier (infrequent transitions).  Bottom: the upper state is more populated; the barrier is small (frequent transitions). The distance between the two potential minima ranges from 200 nm to 700 nm. 
  }
  \label{fig:potentials}
\end{figure}

To analyze our data, we use a variational principle called ``Maximum Caliber'' that purports to predict dynamical properties in the same way that the principle of Maximum Entropy predicts equilibria \cite{Jaynes:macroscopic}.
Caliber has previously been shown to be a simple and useful way to derive the flux distributions in diffusive systems, such as in Fick's Law of particle transport, Fourier's Law of heat transport, and Newton's Viscosity Law of momentum transport \cite{Phillips:Seitaridou}.
The Caliber approach is described in the Appendix. In short, we first enumerate the possible trajectories.  The equivalent of a partition function is then constructed as a sum over weights of the trajectories.  Certain mean values are measured, which then fixes the relative weight factors of the trajectories, resulting in the Caliber prediction for the full distribution function over the trajectories.  Consider the types of trajectories shown in Figure \ref{fig:potentials}.  By \textit{trajectory}, we mean one individual time sequence of events over which the particle transitions back and forth many times between states $A$ and $B$.  We model our events, that is, transitions between states, as taking place at discretized time intervals, $\Delta t$, set by the inverse of the sampling rate. A trajectory has $N$ time steps, so it lasts for a total time $N \Delta t$.  We aim to characterize: (1) various dynamical averages over those trajectories and (2) the probability distribution of the many possible trajectories of the system. 

There are four unknown parameters (plus normalization) called Lagrange multipliers which we determine by measuring four other conjugate quantities, or observables.  These observables are the average number of times a switching event occurs between the same states $\langle N_{aa} \rangle$ and $\langle N_{bb} \rangle$, and between different states $\langle N_{ab} \rangle$ and $\langle N_{ba} \rangle$, in a certain trajectory. The unknown parameters conjugate to these observables are the ``statistical weights'' $\alpha$, $\beta$, $\omega_f$, and $\omega_r$, respectively. $\alpha$ is the statistical weight that, given that the system is in state $A$ at time $t$, it is also in state $A$ at time $t + \Delta t$; $\beta$, for staying in state $B$ at time $t + \Delta t$, given that the system was in $B$ at time $t$; $\omega_{f}$, for switching from $A$ to $B$ in the time interval $\Delta t$; and $\omega_{r}$, for switching from $B$ to $A$ in the time interval $\Delta t$. 

The quantity $Q_d$, the \textit{dynamical partition function}, is the sum over the statistical weights of all the different time trajectories.  For $s = 2^{N}$ trajectories of length $N$ time steps, the dynamical partition function is given by
\begin{equation}
\displaystyle{Q_d = \sum_{i}^{s} \left( \alpha^{N_{aa,i}} \beta^{N_{bb,i}} \omega_f^{N_{ab,i}} \omega_r^{N_{ba,i}}  \right)}, 
\label{eq:dynpartfun}
\end{equation}
and the probability of a particular trajectory labeled $i$ is given by
\begin{equation}
\displaystyle{p_i = Q_d^{-1} \left( \alpha^{N_{aa,i}} \beta^{N_{bb,i}} \omega_f^{N_{ab,i}} \omega_r^{N_{ba,i}} \right )}.
\label{trajprob}
\end{equation}
$N_{ab,i}$ is the number of $A\rightarrow B$ transitions in the $i$th trajectory, for example.

The Caliber strategy provides a way to compute all the higher-order cumulants of the trajectory observables.  To do this, we first obtain the values of the Lagrange multipliers by maximizing $Q_d$ subject to the experimentally observed averages, $\langle N_{aa} \rangle$, $\langle N_{bb} \rangle$, $\langle N_{ab} \rangle$, and $\langle N_{ba} \rangle$.  For instance, $\langle N_{ba} \rangle = \left.\frac{\partial \ln Q_d}{\partial \ln \omega_{r}}\right|_{\alpha,\beta,\omega_f}$ and $\langle N_{ab} \rangle = \left.\frac{\partial \ln Q_d}{\partial \ln \omega_{f}}\right|_{\alpha,\beta,\omega_r}$. Then, taking the second and higher derivatives of $Q_d$ gives the higher cumulants of the observables, such as $\langle N_{ba}^2 \rangle - \langle N_{ba} \rangle^2 = \left.\frac{\partial^2 \ln Q_d}{\partial \left(\ln \omega_{r}\right)^2}\right|_{\alpha,\beta,\omega_f}$. 

There are also other quantities of interest.  Let $N_B$ represent the number of units of time that the system spends in state $B$. Then we have for each individual trajectory $N_B = N_{ab} + N_{bb} + N_{0b}$ and $N_A = N_{aa} + N_{ba} + N_{0a}$ where $N_{0b}$ is $0(1)$ if the trajectory begins in state $A(B)$ and $N_{0a}$ is $1(0)$ if the trajectory begins in state $A(B)$.  If the number of steps is sufficiently large, the contribution from initial conditions can be ignored. Hence the variance for $N_B$ is given by $\left.\langle N_B^2 \rangle - \langle N_B \rangle^2 \simeq \frac{\partial^2  \ln Q_d}{\partial \left(\ln \beta\right)^2}\right|_{\alpha,\omega_{r},\omega_{f}}
+ \left.\frac{\partial^2  \ln Q_d}{\partial\left( \ln \omega_{f}\right)^2}\right|_{\alpha,\beta,\omega_r} + 2\left.\frac{\partial^2  \ln Q_d}{\partial \ln \beta \partial \ln \omega_{f}}\right|_{\alpha,\omega_r}$.

Mixed moments and covariances require mixed derivatives of $Q_d$.  For example, 
\begin{equation}
\left.\frac{\partial^2 \ln Q_d}{\partial \ln \omega_{f} \partial \ln \beta}\right|_{\alpha,\omega_r} = \left.\frac{\partial^2 \ln Q_d}{\partial \ln \beta \partial \ln \omega_f}\right|_{\alpha,\omega_r} \label{onsager1}
\end{equation}
which leads to $\left.\frac{\partial \langle N_{bb} \rangle}{\partial \ln \omega_f}\right|_{\alpha,\beta,\omega_r} = \left.\frac{\partial \langle N_{ab} \rangle}{\partial \ln \beta}\right|_{\alpha,\omega_f,\omega_r} = \langle N_{ab} N_{bb} \rangle - \langle N_{ab} \rangle \langle N_{bb} \rangle$.  Hence, given $Q_d$ all trajectory observables and their fluctuations can be computed.

A simple way to compute $Q_d$ is through the matrix propagator \textbf{G}, 
\begin{equation}
\text{\textbf{G}} = 
\left(
	\begin{array}{cc}
		\alpha & \omega_{r} \\
		\omega_{f} & \beta
  \end{array}
\right)
  \label{transfermatrixeq}
\end{equation}
where each element of ${\textbf{G}}$ represents the ``statistical weight'' of transitioning from some initial state during each time step.  We consider here only stationary processes, for which the statistical weights are time-independent, but the Caliber method itself is not limited to such simple dynamics.  We can express $Q_d = \left(1  \:\:\:\:  1\right) \text{\textbf{G}}^{N-1}\left({a_0 \:\:\:\:  b_0} \right)^{T}$, where $N$ is the number of time steps in the trajectory and $\left({a_0 \:\:\:\:  b_0} \right)^{T}$ denotes the initial state probabilties. Thus all the higher cumulants of the observables are analytically simple in the limit of large $N$ (see Appendix). For non-stationary processes, the $\textbf{G}$ matrix will differ at each time step. 

Functional similarities between microscopic models in statistical mechanics and equations of state in thermodynamics allows assignations of undetermined Lagrange multipliers in the Maximum Entropy formalism to physically realizable quantities, such as $\beta \leftrightarrow T^{-1}$ \cite{Dill:MDF}. We now make similar correspondences between the Caliber-derived ``statistical weights'' with probabilities. The four (exponentiated) Lagrange multipliers $\alpha$, $\beta$, $\omega_f$, and $\omega_r$ in matrix \textbf{G} are reminiscent of a Markov chain propagator.  Thus we choose to assign $\alpha \leftrightarrow P\left(A,t+\Delta t \left.\right| A,t \right)$, $\omega_f \leftrightarrow P\left(B,t+\Delta t \left.\right| A,t \right)$, $\beta \leftrightarrow P\left(B,t+\Delta t \left.\right| B,t \right)$, and $\omega_r \leftrightarrow P\left(A,t+\Delta t \left.\right| B,t \right)$; each is a probability of moving between or among states in time $\Delta t$. Thus $\alpha + \omega_{f} = 1$ and
$ \beta+ \omega_{r} = 1 $ enforce probability conservation.   
\textbf{G} becomes $\left(
	\begin{array}{cc}
		1-\omega_{f} & \omega_{r} \\
		\omega_{f} & 1-\omega_{r}
  \end{array}
\right)$ and the master equation follows immediately. The advantage of the Caliber approach is that it readily provides information about trajectory observables not obviously accesible from master equations.

\begin{figure}[h!]
  \begin{center}
  \includegraphics[width=8cm]{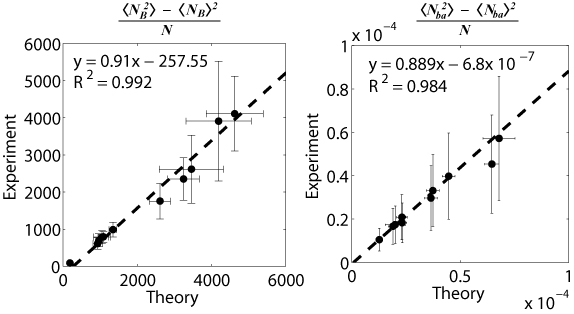}
  \end{center}
  \caption{\textbf{Second cumulant of the trajectory distribution.} The x-axes give the predicted second cumulants from the Caliber approach, based on the known first moments.  The y-axes give the experimental values of the second moments.  Left: variance of $\langle N_{B} \rangle$, right: variance of $\langle N_{ba}\rangle$. The dashed lines are the best linear fits; fitting parameters are inset. Each point represents one experimentally observed trajectory. Trajectories were 30,000 $\Delta t$ units long, and errors were calculated for around 600 trajectories.}
  \label{fig:moment_two}
\end{figure}

We now show tests of the Caliber predictions.  First, given the first-moment averages observed for the trajectories, Caliber predicts the second moments.  Figure \ref{fig:moment_two} demonstrates two predicted second cumulants obtained from two partial derivatives of $Q_d$.  It is in good agreement with the experimental data.  

\begin{figure}[htbp]
  \begin{center}%
  \includegraphics[width=8.5cm]{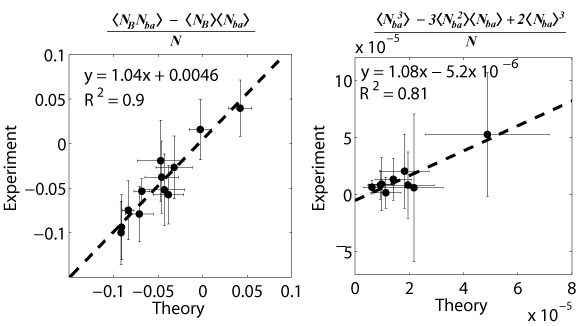}    
  \end{center}
\caption{\textbf{Experiments vs. theory for the covariance and third cumulant.} Left: one covariance quantity.  Right:  The third cumulant of $N_{ba}$. The dashed lines are the best linear fits; fitting parameters are inset.}  
  \label{fig:cross}
\end{figure} 

Figure \ref{fig:cross} compares one experimental third cumulant with the predicted value from Caliber obtained from the measured first moments.  These predictions are also in good agreement with experiments, although, because higher cumulants involve higher derivatives and more data, the scatter is larger than for lower moments.  The first moments are easy to measure with good accuracy from short trajectories, so one virtue of the Caliber approach is that all the higher cumulants, which would require much longer trajectory data, can be predicted from short-trajectory information.

Figure \ref{fig:cross} also shows the quantity $\langle N_{B}N_{ab}\rangle - \langle N_{B}\rangle\langle N_{ab}\rangle$. These covariances, equivalent to mixed moments, give an alternative way to express \textit{reciprocal relationships} resembling the Maxwell relations of thermodynamics and Onsager's reciprocal relations for dynamical processes near equilibrium. In essence, this means that one trajectory observation counts for two: small perturbations on a trajectory are equivalent to observing covariances; thus, without performing additional experiments or recalculating $Q_d$, we know how the system will behave - just looking at the fluctuations is enough.

\begin{figure}[htbp]
  \centering
  \includegraphics[width=8.5cm]{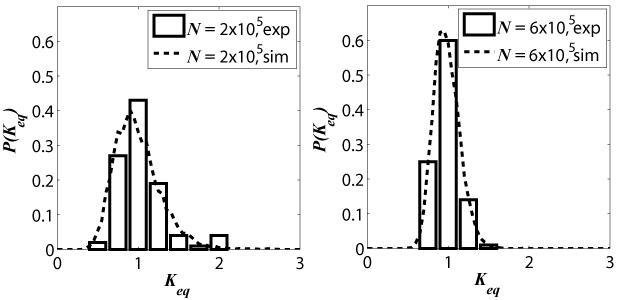}  
  \caption{\textbf{The probability distribution of} $N_{A}/N_{B} = K_{eq}$ \textbf{as a function of time.} We obtain the dashed line from Monte Carlo simulation of $Q_d$ and corresponding columns from experimental data. The distribution of time spent in $A$ versus $B$ is broad for short times (left) (many bad actors) but becomes narrower for increasing trajectory length (right) (fewer bad actors). $N$ denotes the length of each trajectory, each repeated around 100 times. As the length of trajectories increases, the equilibrium constant assumes a delta-function distribution, commensurate with equilibrium assumptions regarding chemical reactions. }
  \label{fig:badactor}
\end{figure}

Using the matrix form of $Q_d$, we can compute the probability distribution of trajectories; we show this for the ratio $N_{A}/N_{B} = K$.  As $t \rightarrow \infty$, this ratio simply becomes equal to the equilibrium constant $K_{eq}$ for the relative populations of the two states $A$ and $B$.  In the small-time limit, this ratio quantity has a distribution of values.  Figure \ref{fig:badactor} shows these distributions for a situation in which the average is $\langle N_{A}/N_{B}\rangle \sim 1 $.  The distribution approaches a $\delta$-function as $t \rightarrow \infty$ and thus $K \rightarrow K_{eq}$.  In diffusion-related problems, small-numbers situations in which particles flow up concentration gradients, rather than down, have been referred to as ``bad actors'' \cite{Phillips:Seitaridou}; the number of bad actors diminishes as trajectories get longer.  Agreement between computation of $Q_d$ and measurements demonstrates that a partition function approach accurately represents the probability distribution of trajectories. 
 
In summary, we have studied a single colloidal particle undergoing a two-state process, $A \leftrightharpoons B$, with stationary rates.  By measuring short trajectories, we obtain first moment observables $\langle N_{bb} \rangle $, $\langle N_{aa} \rangle$, $\langle N_{ba} \rangle$, and $\langle N_{ab} \rangle$.  The variational principle of Maximum Caliber is then used to predict the higher cumulants of the observables as well as the full probability distribution of the trajectories.
Curiously, Maximum Caliber also provides the response function to trajectory perturbation characterized by Maxwell-like relations. Trajectory-based dynamical modeling such as this may be useful in single-molecule and few-molecule science.

We are grateful for the comments of Dave Drabold, Jane Kondev, Keir Neuman, and Dan Gillespie. KD appreciates the support of NIH grant GM 34993 and a UCSF Sandler Blue Sky award. DW acknowledges the support of a NIH UCLA-Caltech MD-PhD fellowship. This work was also supported by the NIH Director's Pioneer award. 
\subsection{\label{sec:appendix_caliber}Appendix}
Suppose we have a set of $ i = 1, 2, \ldots, s$ trajectories.  We aim to determine the probability $p_i$ of each trajectory.  We define an entropy-like quantity called Caliber, ${\cal C}$, over the micro-trajectories (rather than over microstates), subject to dynamical constraints. 
\begin{equation}
{\cal C} = -\sum_i p_i\ln p_i - \lambda\sum_i p_i - \sum_j \lambda_{j} \sum_i p_i N_{j,i}
\end{equation}
where \textit{j} indexes first moment constraints, for instance $\langle N_{aa} \rangle = \sum_i N_{aa_i} p_i$. 

Caliber prescribes that the observed distribution of trajectories will be those $p_i$'s that maximize ${\cal C}$, $\frac{\partial {\cal C}}{\partial p_i} = 0 $.  The corresponding statistical weights are given by $\alpha = \exp(-\lambda_{aa})$, $\beta =  \exp(-\lambda_{bb})$, $\omega_f = \exp(-\lambda_{ab})$, and $\omega_r = \exp(-\lambda_{ba})$, where the $\lambda$'s are the Lagrange multipliers. These Lagrange multipliers can be interpreted as log transition probabilities, in the same spirit of physical interpretation that we assign to the Lagrange multipliers of equilibrium statistical mechanics. 

As the number of time steps increases, using the largest eigenvalue of \textbf{G} as a proxy for $Q_d$ becomes more accurate; it is given by $\xi = \frac{\alpha + \beta + {\sqrt{(\alpha-\beta)^2 + 4\omega_{f}\omega_{r}}}}{2}$. Therefore, for sufficiently long trajectories, we can express all cumulants of the observables analytically in terms of partial derivates of $\xi$.
%
%

\end{document}